\documentclass[final]{siamart251216}

\usepackage{amssymb,amsfonts,amscd,mathrsfs,verbatim}
\usepackage{upgreek}
\usepackage{tikz}

\headers{Two-Dimensional Tomography and Fourier Analysis}{A. Mas, F. Terzioglu, and I. C.F.  Ipsen}

\title{The Role of Fourier Analysis in Two Dimensional Tomography}

\author{Andre T. Mas\thanks{Department of Mathematics, North Carolina State University, Raleigh, NC 27695, USA
(\email{atmas@ncsu.edu}, \email{fterzioglu@ncsu.edu}, \email{ipsen@ncsu.edu}).}
\and Fatma Terzioglu\footnotemark[1]
\and Ilse C.F. Ipsen\footnotemark[1]}

\ifpdf
\hypersetup{
  pdftitle={The Role of Fourier Analysis in Two Dimensional Tomography},
  pdfauthor={Andre T. Mas, Fatma Terzioglu, Ilse C.F. Ipsen}
}
\fi

\begin{document}
\maketitle

\begin{abstract}
We highlight the important role of the Fourier transform in deriving inversion formulas for the integral transforms of tomographic imaging. We demonstrate this principle by deriving inversion formulas for the divergent beam transform and the V-line transform, the latter arising in contemporary models of single-scattering optical tomography.
\end{abstract}



\section{Introduction}\label{intro}

\begin{figure}[t]
    \centering
    \includegraphics[width=0.5\textwidth]{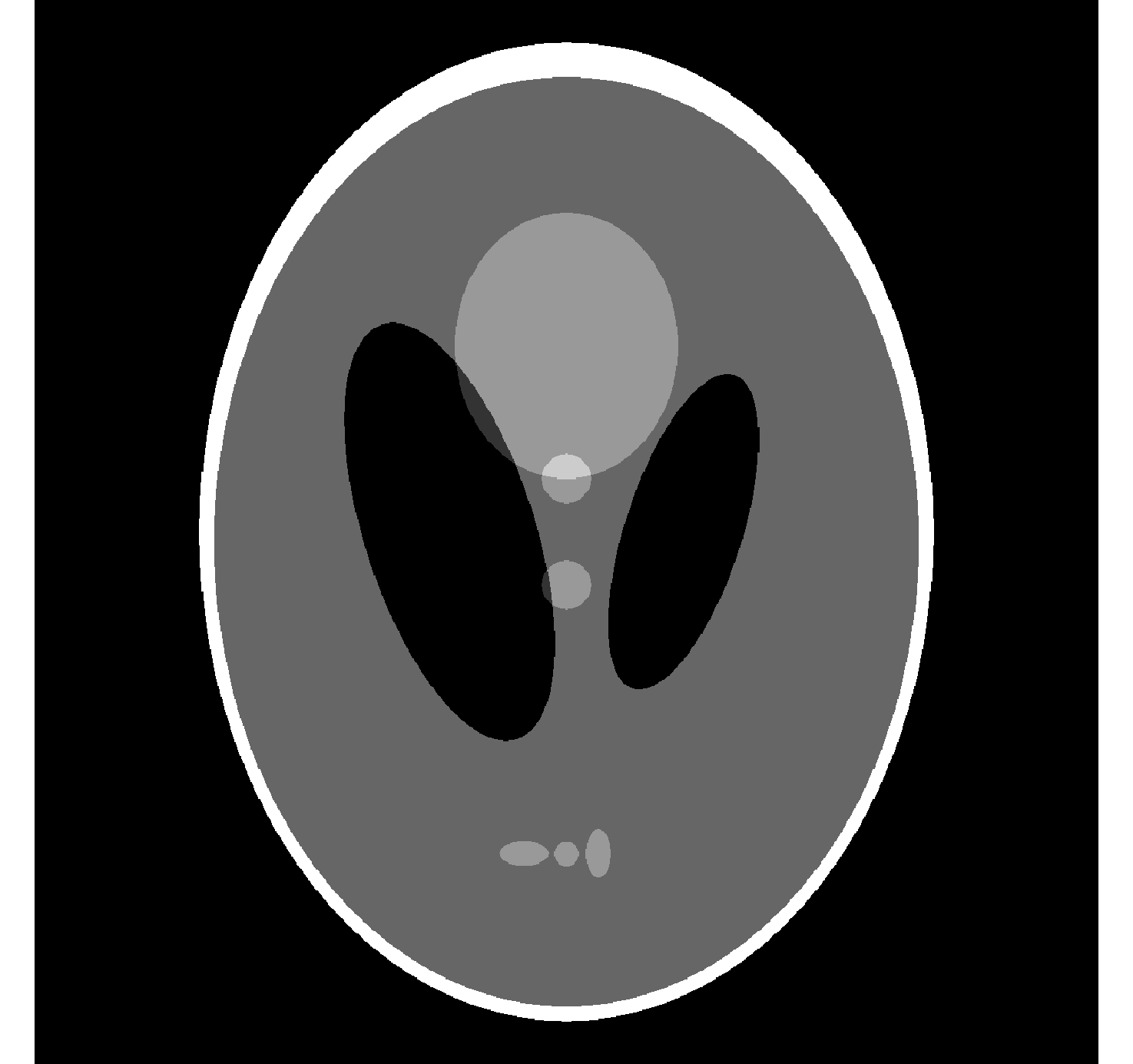}
    \caption{The Shepp-Logan phantom is a common test image created to evaluate tomographic reconstruction
    algorithms. It represents the image a doctor might see when imaging the inside of the human skull. This sample image was generated using the ``phantom'' function in MATLAB.}
    \label{shepp_logan}
\end{figure}

Tomographic imaging is an important part of health and medicine, where the goal is to visualize the interior of the body for diagnostic purposes \cite{sethi_x-rays_2006}. Examples of imaging modalities include x-ray, Magnetic Resonance Imaging (MRI), Positron Emission Tomography (PET), and CT scans. One way to perform tomographic imaging is by emitting particles from a radiative source, which travel through the body and arrive at various detectors. In the context of x-ray imaging, the resulting image that we see at the dentist or the hospital (\cref{shepp_logan}) represents how much intensity the x-ray (or optical ray) lost as it traveled along a path \cite{shepp_reconstructing_1974}. This is measured in terms of attenuation, or reduction of radiative intensity as particles travel through the body. Mathematically, this corresponds to solving a so-called \textit{inverse problem}, where an unknown function is recovered from measurements of its path integrals.

\subsection{Contributions}

The goal of this article is to highlight the important role of the Fourier transform in converting integral transform equations into algebraic equations, from which inversion formulas may be read off directly. We demonstrate this principle by providing the derivation of inversion formulas for the divergent beam and V-line transforms respectively.

\subsection{Outline}

The remainder of \cref{intro} consists of a brief overview of the general principles of two-dimensional tomography using the divergent beam transform as a motivating example. In \cref{ftsec} we motivate the usage of the Fourier transform in the context of tomography, and we apply it to the problem of inverting the divergent beam transform. In \cref{vline_sec}, we show how the techniques introduced previously can be used to derive inversion formulas for more complicated integral transforms. In terms of physical problems of interest, this provides a tool for handling more complicated particle scattering patterns. This presentation is meant to be understandable for a broad audience, with the only prerequisite knowledge being basic familiarity with the Fourier transform and its analytic properties.

\subsection{The General Problem of Medical Tomography}

The grayscale value displayed in each pixel of a tomographic image is that of the so-called \textit{attenuation coefficient} at each point along the travel path of the particles. The attenuation coefficient is a scalar-valued function that measures the amount of radiative intensity lost due to absorption (neglecting scattering). This is the key quantity that practitioners wish to visualize, as it allows them to identify tumorous tissues. However, the values of the attenuation coefficient at each point along the path are not what the detectors actually measure. The detectors measure the \textit{total attenuation} along the travel path of a particle as it moves through the body. Total attenuation is given by the integral of the attenuation coefficient along the travel path. This relationship between detector measurements and the attenuation coefficient forms the basis for many contemporary tomography models, allowing for the creation of the image that we see on a doctor's computer screen \cite{natterer_mathematics_1986}.

Mathematically, the general problem of (two-dimensional) tomography is as follows. Let $\Gamma$ denote the set of ``allowable'' travel paths for the photons, and $f : \mathbb{R}^2 \to \mathbb{R}$ be a bounded function that represents the attenuation coefficient. This function is the quantity that we wish to recover. It is customary to assume that $f$ is smooth to simplify the resulting analysis. In practice, however, the attenuation coefficient often exhibits jump discontinuities due to the presence of different materials along the travel path -- each of which may attenuate radiation differently. For example, bone causes higher levels of attenuation than water since bone has a higher density. The data measured at each detector may be modeled as the output of a new function $\mathcal{T}f : \Gamma \to \mathbb{R}$ that integrates $f$ along a travel path $\gamma \in \Gamma$:
\[
\mathcal{T}f(\gamma) := \int_{\gamma} f(y)\,dy.
\]
This integral represents total loss in intensity (or total attenuation) along the travel path $\gamma$, depending on the attenuation coefficient $f$. As the notation suggests, $\mathcal{T}f$ is a linear operator~$\mathcal{T}$ applied to $f$. Such operators are referred to as \textit{integral transforms}, and are the central objects of interest in the field of \textit{integral geometry} \cite{helgason_integral_2011}. Recovery of~$f$ then amounts to finding a suitable inverse for the integral transform~$\mathcal{T}$. This is the central question behind the mathematics of tomography.

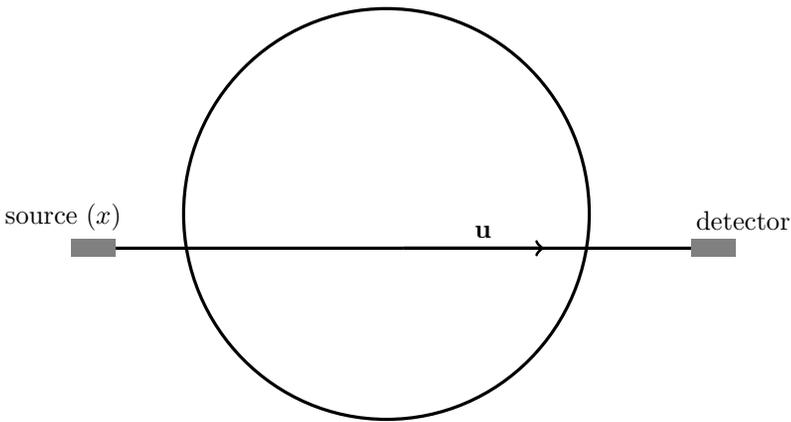
\begin{figure}[ht]
\centering
\begin{tikzpicture}[scale=1.5]
\coordinate (x) at (0,0);
\coordinate (circlecenter) at (-0.15,0.3);
\draw[very thick, black] (circlecenter) circle (1.8);

\draw[->,black, very thick] (x) -- (180:2.75)
    node[above, black, xshift=-4mm, yshift=1mm] {source $(x)$}
    node[rectangle, fill=gray] {\,\,\,\,\,\,};

\draw[->,black, very thick] (x) -- (0:2.75)
    node[midway, above, text=black, xshift=-10mm] {$\mathbf{u}$}
    node[above, black, xshift=4mm, yshift=1mm] {detector}
    node[rectangle, fill=gray] {\,\,\,\,\,\,};

  \draw[->,black, very thick] (x) -- (0:1.25);

\end{tikzpicture}
\caption{Imaging geometry for the divergent beam transform. Particles are emitted from
  a source located at point $x$, travel through the object along a line determined by direction
  $\mathbf{u}$, and arrive at a detector.}
\label{dbtgeom}
\end{figure}

Different integral transforms arise depending on the set of allowable travel paths $\Gamma$, or more generally the sets upon which integration takes place. For example, travel paths consisting of lines or rays are used in models of x-ray imaging. In three dimensions, integration over planes models MRI measurements, and integration over cones models Compton camera measurements \cite{terzioglu_inversion_2015}. For now we choose to focus on travel paths consisting of rays -- an example of which may be seen in \cref{dbtgeom}. The resulting integral transform is known as the \textit{divergent beam transform}. In two dimensions this transform is often referred to as the \textit{fan-beam transform} \cite{fanbeam_fourier}, and in three dimensions the term \textit{cone beam transform} is common in the literature. We will continue to use the term ``divergent beam'' consistent with literature addressing the V-line transform, which will be discussed in forthcoming sections.

In the context of the divergent beam transform, rays are parameterized in terms of a source point $x \in \mathbb{R}^2$ and a direction vector $\mathbf{u} \in \mathbb{S}^1$, where $\mathbb{S}^1$ denotes the unit circle in $\mathbb{R}^2$. The divergent beam transform applied to a function $f$ then takes the form:
\begin{equation}\label{DBT_Def}
    \mathcal{X} f(x,\mathbf{u}) := \int_{0}^\infty f(x+t\mathbf{u}) \,dt.
\end{equation}

In practice, the point $x$ represents the location of a radiative source, and the vector $\mathbf{u}$ determines a direction pointing towards a detector. The physical distance from the source to the detector is always finite, so it is possible to instead write the definition in~\cref{DBT_Def} with a finite upper limit of integration. That being said, an infinite limit of integration has nice analytic properties which are necessary for the manipulations that will follow in \cref{ftsec}. Since $f$ is compactly supported in practice (the patient is not infinitely large!), both limits of integration produce the same integral: the integrand vanishes beyond its support, so extending the upper limit to infinity does not change the value of the integral.

As mentioned previously, the general integral transforms that arise in tomography are \textit{linear operators}. That is, for any real number $c$ and functions $f$ and $g$:
\begin{equation}
    \mathcal{T}[cf+g](\cdot) = c\mathcal{T}f(\cdot) + \mathcal{T}g(\cdot).
\end{equation}

Recall from linear algebra that a necessary condition for a matrix to be invertible is that the matrix is square. A similar phenomenon exists in the context of linear integral transforms. Given that $f : \mathbb{R}^2 \to \mathbb{R}$ depends on a point in $\mathbb{R}^2$ which has \textit{two} degrees of freedom, one would hope that the corresponding function $\mathcal{T}f$ also has two degrees of freedom.

In the case of the divergent beam transform, $\mathcal{X} f(x,\mathbf{u})$ depends on a point $x \in \mathbb{R}^2$ (two parameters) and a direction vector $\mathbf{u} \in \mathbb{S}^1$ (one angular parameter), so it has $2+1 = 3$ degrees of freedom. The problem of inverting the divergent beam transform is therefore \textit{overdetermined} in the sense of linear algebra.

In practice, one restricts the location of sources and detectors based on the imaging modality, which corresponds to fixing several of the parameters in each scenario. In the ideal case with x-ray imaging, enough parameters are fixed so that the corresponding function $\mathcal{X}f$ depends on only two parameters. For illustrative purposes, we fix the directional input $\mathbf{u}$ while letting the point $x$ contribute the two relevant degrees of freedom. We denote this fixed-direction divergent beam transform by $\mathcal{X}_{\mathbf{u}}$, so that $\mathcal{X}_{\mathbf{u}}f : \mathbb{R}^2 \to \mathbb{R}$ is defined by:
\begin{equation}
    \mathcal{X}_{\mathbf{u}} f(x) := \int_{0}^\infty f(x+t\mathbf{u}) \,dt.
\end{equation}

Inverting the divergent beam transform in this case \cite[page~50]{sherson_results_2015} amounts to an application of the directional derivative operator $\mathcal{D}_\mathbf{u} := \mathbf{u} \cdot \nabla$. That is,
\begin{equation}\label{ftoc1}
    -\mathcal{D}_\mathbf{u}\mathcal{X}_{\mathbf{u}} f(x) = f(x).
\end{equation}

\begin{figure}[ht]
\centering
\begin{tikzpicture}[scale=1]

\begin{scope}[shift={(-2.6,0)}]
\coordinate (x) at (0,0);
\draw[->,black, very thick] (x) -- (180:2.2);
\draw[->,black, very thick] (x) -- (45:2.4);
\fill[black] (x) circle (2pt) node[below left] {~};
\end{scope}

\begin{scope}[shift={(2.6,0)}]
\coordinate (x) at (0,0);
\foreach \ang in {90, 18, -54, -126, 162}
{
  \draw[->,black, very thick] (x) -- (\ang:2.4);
}
\fill[black] (x) circle (2pt) node[below left] {~};
\end{scope}
\end{tikzpicture}
\caption{Example of broken-ray (left) and star (right) geometries, which arise in
  models of single scattering tomography.}
\label{V-line_and_star_geometries}
\end{figure}
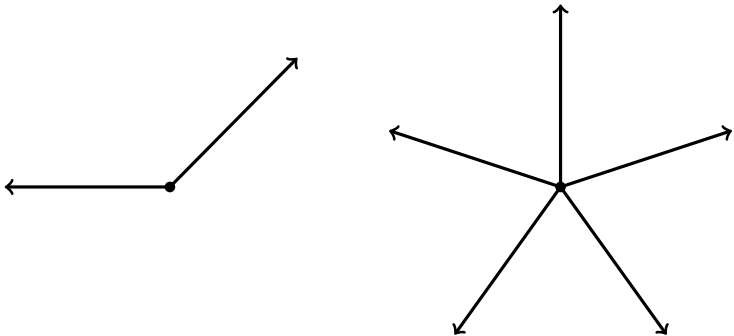

As travel paths become more complicated, such as the cones of Compton camera imaging \cite{terzioglu_inversion_2015} or the ``broken-rays'' and ``stars'' of single-scattering tomography \cite{ambartsoumian_generalized_2023} illustrated in \cref{V-line_and_star_geometries}, so too do the techniques necessary for inverting the integral transform of interest. This results from the fact that the fundamental theorem of calculus does not generalize as readily to integration over arbitrary surfaces. The idea is to use another tool that interacts nicely with derivatives and integration, simplifying the procedure of ``undoing'' these transforms in a more general setting.

\section{The Fourier Transform}\label{ftsec}

The Fourier transform is an important tool in applied mathematics that finds broad usage in areas such as image and signal processing~\cite{shepp_fourier_1974}. We take a brief intermission to review useful properties of the Fourier transform, then return to the problem of inverting the divergent beam transform using this tool. For an integrable function $f : \mathbb{R}^2 \to \mathbb{R}$, the Fourier transform is a linear operator defined by:
\begin{equation}\label{ftdef}
    \widehat{f}(\xi) = \mathscr{F} \{f\}(\xi) := \int_{\mathbb{R}^2} f(y)e^{-i2\pi (y \cdot \xi)}\,dy,
\end{equation}
where $i$ denotes the imaginary unit, and $(y \cdot \xi)$ denotes the dot product of the points $y := (y_1,y_2)$ and $\xi := (\xi_1,\xi_2)$ in $\mathbb{R}^2$. Importantly, for a sufficiently nice function the Fourier transform is readily invertible via the formula:
\begin{equation}\label{ftinvdef}
    f(y) = \mathscr{F}^{-1} \left\{\widehat{f}\right\}(y) := \int_{\mathbb{R}^2} \widehat{f}(\xi)e^{i2\pi (\xi \cdot y)}\,d\xi.
\end{equation}

One property of the Fourier transform that makes it useful in the analysis of differential equations and integral transforms is the way in which it interacts with derivatives and integrals. Taking a directional derivative in the direction $\mathbf{u} = (u_1,u_2)$ corresponds, in the Fourier domain, to multiplication by $i2\pi(\mathbf{u} \cdot \xi)$. Indeed,
\begin{align}\label{ftderiv}
    \mathscr{F} \{\mathcal{D}_{\mathbf{u}}f\}(\xi) &= u_1\mathscr{F} \{\partial_{y_1}f\}(\xi) + u_2 \mathscr{F} \{\partial_{y_2}f\}(\xi) \\
    &= i2\pi u_1\xi_1 \int_{\mathbb{R}^2} f(y)e^{-i2\pi(y\cdot\xi)}\,dy \nonumber \\
    &\quad + i2\pi u_2\xi_2 \int_{\mathbb{R}^2} f(y)e^{-i2\pi(y\cdot\xi)}\,dy \nonumber \\
    &= i2\pi(\mathbf{u} \cdot \xi) \widehat{f}(\xi). \nonumber
\end{align}

\subsection{The Key Idea with the Fourier Transform}

Conversely to equation \cref{ftderiv}, integration in the direction $\mathbf{u}$ (i.e., applying $\mathcal{X}_{\mathbf{u}}$) corresponds to division by $i2\pi(\mathbf{u}\cdot\xi)$ in the Fourier domain if $\mathbf{u}\cdot\xi \neq 0$, as we shall see. Consequently, ``undoing'' an integral transform (i.e., finding an inversion formula) amounts to multiplying by an appropriate factor in the Fourier domain and then applying the inverse Fourier transform.

These algebraic properties make the Fourier transform an exceptionally powerful tool for inverting integral transforms: rather than working with complicated operator equations, one can reduce them to pointwise algebraic equations in the Fourier domain, solve for $\widehat{f}(\xi)$, and recover $f$ by applying the inverse Fourier transform.

\subsection{Fourier-Inversion of the Divergent Beam Transform}\label{fidbt}

Let us revisit the problem of inverting the divergent beam transform by taking the Fourier transform of $\mathcal{X}_{\mathbf{u}} f$. Recall from the discussion above that the strategy is to express $\mathscr{F}\{\mathcal{X}_\mathbf{u} f\}(\xi)$ as $\widehat{f}(\xi)$ times some factor depending only on $\xi$; once this is done, one can solve algebraically for $\widehat{f}(\xi)$ and recover $f$ via the inverse Fourier transform. We proceed by swapping the order of integration and applying the substitution $w = y + t\mathbf{u}$:
\begin{align*}
    \mathscr{F}\{\mathcal{X}_{\mathbf{u}} f\}(\xi)
      &= \int_{\mathbb{R}^2}\int_0^\infty f (y+t\mathbf{u}) e^{-i2\pi(y \cdot \xi)} \,dt\,dy \\
      &= \int_0^\infty \int_{\mathbb{R}^2} f(w) e^{-i2\pi(w \cdot \xi)}
         e^{i2\pi t(\mathbf{u} \cdot \xi)} \,dw\,dt \\
      &= \widehat{f}(\xi) \int_0^\infty e^{i2\pi t(\mathbf{u} \cdot \xi)} \,dt \\
      &= \widehat{f}(\xi) \left(\frac{-1}{i 2 \pi (\mathbf{u} \cdot \xi)}
         + \frac{1}{2}\delta(\mathbf{u} \cdot \xi)\right).
\end{align*}

In the above computation, $\delta(\cdot)$ denotes the Dirac delta distribution, defined by the relations $\delta(x) = 0$ if $x \neq 0$ and:

\[
\int_{-\infty}^\infty g(x)\delta(x)\,dx = g(0).
\]

The full details of computing the integral

\begin{equation}\label{appendix_eq}
\int_0^\infty e^{i2\pi t(\mathbf{u} \cdot \xi)} \,dt
  = \frac{-1}{i 2 \pi (\mathbf{u} \cdot \xi)} + \frac{1}{2}\delta(\mathbf{u} \cdot \xi),
\end{equation}

may be found in the appendix, and involves computing the Fourier transform of the Heaviside or ``unit-step function.'' Through the above series of computations, we arrive at the key formula:
\begin{equation}\label{ftdbt}
\mathscr{F}\{\mathcal{X}_{\mathbf{u}} f\}(\xi) = \widehat{f}(\xi)
  \left(\frac{-1}{i 2 \pi (\mathbf{u} \cdot \xi)}
  + \frac{1}{2}\delta(\mathbf{u} \cdot \xi)\right).
\end{equation}

Multiplying both sides by $-i 2 \pi (\mathbf{u} \cdot \xi)$ gives:
\begin{equation}\label{ft_inv_dbt}
    -i 2 \pi (\mathbf{u} \cdot \xi) \mathscr{F}\{\mathcal{X}_{\mathbf{u}} f\}(\xi)
      = \widehat{f}(\xi) - v_{\mathbf{u}}(\xi) \widehat{f}(\xi),
\end{equation}

where $v_{\mathbf{u}}(\xi) = i \pi (\mathbf{u} \cdot \xi) \delta(\mathbf{u} \cdot \xi)$. We then apply the inverse Fourier transform to both sides, which yields $f$ on the right hand side. We note that for the second term on the right:

\begin{align}\label{deltavanish}
    \mathscr{F}^{-1}\{v_{\mathbf{u}} \widehat{f}\}(x) &= \int_{\mathbb{R}^2} v_{\mathbf{u}}(\xi) \widehat{f}(\xi) e^{i2\pi(\xi \cdot x)} \,d\xi \\
    &= \int_{\mathbf{u} \cdot \xi = 0} i \pi (\mathbf{u} \cdot \xi) \delta(\mathbf{u} \cdot \xi) \widehat{f}(\xi) e^{i2\pi(\xi \cdot x)} \,d\xi \nonumber \\
    &\quad + \int_{\mathbf{u} \cdot \xi \neq 0} i \pi (\mathbf{u} \cdot \xi) \delta(\mathbf{u} \cdot \xi) \widehat{f}(\xi) e^{i2\pi(\xi \cdot x)} \nonumber \,d\xi .
\end{align}

In the second equality, the first integral vanishes as the integrand is zero, and the second integral vanishes by definition of the Dirac distribution. Applying $\mathscr{F}^{-1}$ to both sides of \cref{ft_inv_dbt} and using $\mathscr{F}^{-1}\{v_{\mathbf{u}} \widehat{f}\}(x) = 0$ as well as equation \cref{ftderiv}, one again obtains:
\begin{equation*}
    -\mathcal{D}_\mathbf{u}\mathcal{X}_{\mathbf{u}} f(x) = f(x),
\end{equation*}
as shown previously in equation \cref{ftoc1}. At first glance, the above process seems merely like a more complicated procedure for obtaining this inversion formula. However, we will now move to the problem of inverting another integral transform that arises in two-dimensional tomography, where the Fourier transform will become even more useful.

\section{The V-line Transform}\label{vline_sec}

\begin{figure}[ht]
\centering
\begin{tikzpicture}[scale=1.5]
\coordinate (x) at (0,0);
\coordinate (circlecenter) at (-0.15,0.3);
\draw[very thick, black] (circlecenter) circle (1.8);
\draw[->,black, very thick] (x) -- (180:2.75) 
node[midway, above, xshift=4mm, text=black] {$\mathbf{u}_{1}$} 
node[above, black, xshift=-4mm, yshift=1mm] {source} 
node[rectangle, fill=gray] {\,\,\,\,\,\,};
\draw[->,black, very thick] (x) -- (45:3) 
node[midway, above, xshift=-8mm, yshift=-6mm, text=black] {$\mathbf{u}_{2}$} 
node[below right, black, yshift=-0.25mm] {detector} 
node[rectangle, fill=gray, rotate=45] {\,\,\,\,\,\,};
\fill[black] (x) circle (2pt) node[below left, yshift=-0.5mm] {\textcolor{black}{$x$}};
\end{tikzpicture}
\caption{Imaging geometry for the V-line transform. A particle travels from a source along
  direction $\mathbf{u}_1$, scatters at the point $x$, and then travels toward a detector
  along direction $\mathbf{u}_2$. The two rays form a ``V'' shape at the scattering vertex.}
\label{vline_geom}
\end{figure}
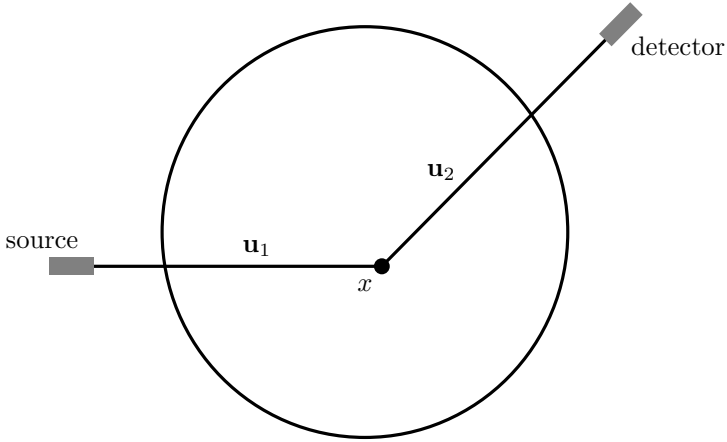

In the last two decades, considerable progress has been made towards models of x-ray and optical tomography that include scattering effects \cite{desmal_limited-view_2019, florescu_single-scattering_2010, florescu_inversion_2011, florescu_single-scattering_2009, greenberg_design_2021, krylov_inversion_2015, walker_broken_2019}. In particular, under the assumption that the optical thickness of the imaged object is small (specifically, that the so-called ``mean-free path'' of particles is comparable to the size of the object), particles may be assumed to scatter at most once as they travel through the region of interest. This is sometimes referred to as the \textit{single-scattering approximation}. In Sections~1 and~2, particles are assumed to travel in straight lines without scattering at all. In this section we relax that assumption to allow exactly one scattering event inside the medium. The single-scattering approximation results in measured data corresponding to attenuation along a V-shaped path. These data are modeled using the V-line (or broken ray) transform, which consists of two divergent beam transforms added together. In this setting, the two rays $\mathbf{u}_1$ and $\mathbf{u}_2$ point in the directions of a source and detector, respectively, and the point $x$ represents the location of the scattering event rather than a radiative source as seen in \cref{vline_geom}.

A closed-form expression for the V-line transform is:
\begin{equation}
    \mathcal{V} f (x,\mathbf{u}_1,\mathbf{u}_2)
      := \mathcal{X} f (x,\mathbf{u}_1) + \mathcal{X} f (x,\mathbf{u}_2).
\end{equation}

The V-line transform has been studied extensively since its introduction \cite{ambartsoumian_inversion_2012, ambartsoumian_inversion_2021, ambartsoumian_v-line_2019, gouia-zarrad_exact_2014, katsevich_broken_2013}. These works employ a variety of methods: geometric techniques based on the symmetry of the V-line \cite{ambartsoumian_inversion_2012}, algebraic \cite{katsevich_broken_2013} and microlocal approaches \cite{sherson_results_2015}, and explicit inversion using special function representations \cite{gouia-zarrad_exact_2014, haltmeier_inversion_2017}. The apparent variety and complexity of methods is due primarily to the fact that the general inverse problem is significantly more involved than what we present below. Here, we focus on the simpler fixed-direction case, which nonetheless captures the key ideas and is of practical relevance.

We focus on inverting the V-line transform with fixed directional inputs. Fixing both directions $\mathbf{u}_1$ and $\mathbf{u}_2$ is a natural assumption in practice, since in a physical scanner the angles at which particles are emitted and detected are typically held constant while the source and detectors are moved around the object \cite{kuchment_inversion_2017}. This results in a function $\mathcal{V}_{\mathbf{u}_1,\mathbf{u}_2}f : \mathbb{R}^2 \to \mathbb{R}$ that depends only on the spatial variable $x$, making the Fourier transform directly applicable. We denote this fixed-direction V-line transform by:
\begin{equation}
    \mathcal{V}_{\mathbf{u}_1,\mathbf{u}_2} f (x)
      := \mathcal{X}_{\mathbf{u}_1} f (x) + \mathcal{X}_{\mathbf{u}_2} f (x).
\end{equation}

Notice that the V-line transform is again linear in $f$. Using \cref{ftdbt} alongside the linearity of the Fourier transform, we derive an inversion formula for the V-line transform following the approach first proposed in \cite{walker_broken_2019}.

\subsection{Fourier-Inversion of the V-line Transform}

The strategy mirrors that of section \ref{fidbt}: we take the Fourier transform of $\mathcal{V}_{\mathbf{u}_1,\mathbf{u}_2} f$, use linearity and \cref{ftdbt} to express the result as $\widehat{f}(\xi)$ times an explicit factor, and then solve algebraically for $\widehat{f}(\xi)$ before inverting the Fourier transform:

\begin{align*}
    \mathscr{F}\{\mathcal{V}_{\mathbf{u}_1,\mathbf{u}_2} f\}(\xi)
      &= \mathscr{F}\{\mathcal{X}_{\mathbf{u}_1} f\}(\xi)
         + \mathscr{F}\{\mathcal{X}_{\mathbf{u}_2} f\}(\xi) \\
      &= \widehat{f}(\xi) \Bigg(
         \frac{-1}{i 2 \pi (\mathbf{u}_1 \cdot \xi)}
         + \frac{1}{2}\delta(\mathbf{u}_1 \cdot \xi) \\
      &\quad\quad\quad\quad
         + \frac{-1}{i 2 \pi (\mathbf{u}_2 \cdot \xi)}
         + \frac{1}{2}\delta(\mathbf{u}_2 \cdot \xi)\Bigg).
\end{align*}

To isolate $\widehat{f}(\xi)$, we need to simplify the second factor on the right-hand side. Combining the two rational terms over a common denominator and using bilinearity of the dot product to write $-(\mathbf{u}_1\cdot\xi) - (\mathbf{u}_2\cdot\xi) = -(\mathbf{u}_1 + \mathbf{u}_2)\cdot\xi$:
\begin{equation}\label{bigvlteq}
    \mathscr{F}\{\mathcal{V}_{\mathbf{u}_1,\mathbf{u}_2} f\}(\xi)
      = \widehat{f}(\xi) \left(
        \frac{-(\mathbf{u}_1 + \mathbf{u}_2)\cdot\xi}
             {i2\pi(\mathbf{u}_1\cdot\xi)(\mathbf{u}_2\cdot\xi)}
        + \frac{1}{2}\delta(\mathbf{u}_1 \cdot \xi)
        + \frac{1}{2}\delta(\mathbf{u}_2 \cdot \xi)\right).
\end{equation}

The goal now is to recognize the factor behind $\widehat{f}(\xi)$ in terms of the Fourier-domain operations we know: multiplication by $i2\pi(\mathbf{u}\cdot\xi)$ corresponds to directional differentiation, and division by $i2\pi(\mathbf{u}\cdot\xi)$ corresponds to integration. To make this correspondence explicit, we multiply the numerator and denominator of the first fraction by $i2\pi$:
\begin{equation}\label{VLTimp}
    \mathscr{F}\{\mathcal{V}_{\mathbf{u}_1,\mathbf{u}_2} f\}(\xi)
      = \widehat{f}(\xi) \left(
        \frac{-i2\pi(\mathbf{u}_1 + \mathbf{u}_2) \cdot \xi}
             {(i2\pi)^2(\mathbf{u}_1\cdot\xi)(\mathbf{u}_2\cdot\xi)}
        + \frac{1}{2}\delta(\mathbf{u}_1 \cdot \xi)
        + \frac{1}{2}\delta(\mathbf{u}_2 \cdot \xi)\right).
\end{equation}

We now multiply both sides of \cref{VLTimp} by the inverse of the first factor inside the parentheses, i.e., by $(i2\pi)^2(\mathbf{u}_1\cdot\xi)(\mathbf{u}_2\cdot\xi) / (-i2\pi(\mathbf{u}_1+\mathbf{u}_2)\cdot\xi)$ for $(\mathbf{u}_1+\mathbf{u}_2)\cdot\xi\neq 0$, isolating $\widehat{f}(\xi)$ up to an additive term $v_{\mathbf{u}_1,\mathbf{u}_2}(\xi)$ containing the Dirac distributions:
\begin{equation}\label{VLTimp2}
    \frac{(i2\pi)^2(\mathbf{u}_1\cdot\xi)(\mathbf{u}_2\cdot\xi)}
         {-i2\pi(\mathbf{u}_1 + \mathbf{u}_2) \cdot \xi}
    \mathscr{F}\{\mathcal{V}_{\mathbf{u}_1,\mathbf{u}_2} f\}(\xi)
      = \widehat{f}(\xi) \left(
        1 + v_{\mathbf{u}_1,\mathbf{u}_2}(\xi)\right).
\end{equation}

The term $v_{\mathbf{u}_1,\mathbf{u}_2}(\xi) \widehat{f}(\xi)$ on the right vanishes under the inverse Fourier transform by the same argument as in \cref{deltavanish}. Since $\mathbf{u}_1 + \mathbf{u}_2$ is not necessarily unit-length, we introduce the unit vector $\mathbf{v} := (\mathbf{u}_1 + \mathbf{u}_2)/\|\mathbf{u}_1 + \mathbf{u}_2\|$, $\mathbf{u}_1 \neq -\mathbf{u}_2$, to separate the scalar factor $\|\mathbf{u}_1 + \mathbf{u}_2\|$ from the directional information. Rewriting \cref{VLTimp2} accordingly:
\begin{equation}\label{VLTimp3}
    \frac{(i2\pi)^2(\mathbf{u}_1\cdot\xi)(\mathbf{u}_2\cdot\xi)}
         {-i2\pi\|\mathbf{u}_1 + \mathbf{u}_2\|\, \mathbf{v} \cdot \xi}
    \mathscr{F}\{\mathcal{V}_{\mathbf{u}_1,\mathbf{u}_2} f\}(\xi)
      = \widehat{f}(\xi) \left(
        1 + v_{\mathbf{u}_1,\mathbf{u}_2}(\xi)\right).
\end{equation}

Reading off the left-hand side of \cref{VLTimp3}: the two multiplicative factors $(i2\pi)(\mathbf{u}_1\cdot\xi)$ and $(i2\pi)(\mathbf{u}_2\cdot\xi)$ in the numerator correspond to directional derivatives in $\mathbf{u}_1$ and $\mathbf{u}_2$ respectively, while the factor $-i2\pi(\mathbf{v}\cdot\xi)$ in the denominator corresponds to integration in the direction $\mathbf{v}$. The scalar $\|\mathbf{u}_1+\mathbf{u}_2\|$ passes through as a constant. Applying $\mathscr{F}^{-1}$ to both sides and discarding the vanishing term yields the inversion formula:
\begin{equation}\label{VLTinv}
    f(x) = \frac{1}{\|\mathbf{u}_1 + \mathbf{u}_2\|}
             \mathcal{D}_{\mathbf{u}_1}\mathcal{D}_{\mathbf{u}_2}
             \int_0^\infty \mathcal{V}_{\mathbf{u}_1,\mathbf{u}_2} f\left(x+t\mathbf{v}\right)\,dt.
\end{equation}

Using the shorthand $c_{\mathbf{u}_1,\mathbf{u}_2} = 1/\|\mathbf{u}_1 + \mathbf{u}_2\|$, this may be written more succinctly as:
\begin{equation}\label{VLTinv2}
    f(x) = c_{\mathbf{u}_1,\mathbf{u}_2}
             \mathcal{D}_{\mathbf{u}_1}\mathcal{D}_{\mathbf{u}_2}
             \mathcal{X}_\mathbf{v} (\mathcal{V}_{\mathbf{u}_1,\mathbf{u}_2} f)(x), \quad \mathbf{v} = \frac{\mathbf{u}_1 + \mathbf{u}_2}{\|\mathbf{u}_1 + \mathbf{u}_2\|}.
\end{equation}

This operator-theoretic form is worth interpreting carefully as it explains how to recover $f$ at a point $x$ given the V-line data $\mathcal{V}_{\mathbf{u}_1,\mathbf{u}_2}f$. Namely, one integrates along direction $\mathbf{v}$ via $\mathcal{X}_\mathbf{v}$, applies the two derivatives $\mathcal{D}_{\mathbf{u}_1}$ and $\mathcal{D}_{\mathbf{u}_2}$, and scales by $c_{\mathbf{u}_1,\mathbf{u}_2}$. Reconstruction plots and numerical results for this inversion formula may be found in \cite{walker_broken_2019}. Each step corresponds precisely to an algebraic operation carried out in the Fourier domain.

\subsection{A General Framework}

The effectiveness of the Fourier transform in deriving the inversion formulas above suggests a unified framework for understanding and inverting integral transforms arising in two-dimensional tomography. By fixing input parameters until only two free variables remain, it is possible to use the properties of the Fourier transform to convert the inverse problem into a pointwise algebraic equation in the Fourier domain. In fact, this principle underlies the derivation of inversion formulas for many integral transforms beyond the two-dimensional setting. 

For example, one may use the Fourier transform to derive the ``filtered back-projection'' formula used to invert the \textit{Radon transform} in models of x-ray CT imaging \cite{natterer_mathematics_1986}. Similarly, fixing directional inputs yields an inversion formula for the star transform \cite{ambartsoumian_inversion_2021, zhao_inversion_2014}, which also arises in two-dimensional single scattering optical tomography. While geometric approaches \cite{ambartsoumian_inversion_2021} have been used to derive inversion formulas for the star transform, these techniques do not generalize as readily to other integral transforms.

In practical numerical implementations, the Fast Fourier Transform (FFT) makes the Fourier-based inversion approach computationally attractive. For instance, the aforementioned filtered back-projection formula is implemented via the FFT, and constitutes the standard reconstruction algorithm in modern CT scanners \cite{natterer_mathematics_1986}. For the V-line and broken-ray transforms, FFT-based implementations have likewise been explored \cite{ambartsoumian_numerical_2024}, and the efficiency of the Fast Fourier Transform makes such algorithms feasible for large-scale imaging problems.

\subsection{Current Research Directions and Further Reading}

Inverting integral transforms that arise from physical imaging models is a broad and active research area \cite{desmal_limited-view_2019, greenberg_design_2021}. New imaging modalities continue to motivate new transforms and new inverse problems. The presence of noise, incomplete data, or physical complications such as attenuation and scattering present further challenges in the implementation of the inversion algorithms \cite{nguyen_sampling_2021, nguyen_how_2015, nguyen_scattered_2011}. The interplay between the continuous mathematical theory (existence and uniqueness of inversion formulas, stability of reconstruction) and discrete numerical implementation (efficient algorithms, artifacts, convergence) keeps this field lively across both pure and applied mathematics.

Several specific directions have received considerable recent attention. One such active direction concerns \textit{attenuated} integral transforms, where the integrand includes an additional weight accounting for the particle's loss of intensity along its travel path even before scattering. Inversion formulas for the attenuated V-line and broken-ray transforms have been derived in \cite{haltmeier_inversion_2017, krylov_inversion_2015}, and remain an active topic.

A second direction involves extending these ideas to \textit{vector tomography}, where one attempts to recover a vector field rather than a scalar function. Generalized V-line transforms on vector fields and their numerical inversion have been studied in \cite{ambartsoumian_generalized_2020, ambartsoumian_numerical_2024}.

A third direction concerns \textit{limited-data} problems, where measurements are only available over a restricted range of angles or detector positions. Artifacts arising in such settings have been characterized using microlocal analysis \cite{frikel_characterization_2013, nguyen_how_2015}, a more advanced Fourier-analytic tool that tracks the ``direction'' of singularities in both physical and frequency space \cite{brouder_smooth_2014}. Microlocal analysis of the V-line transform is discussed extensively in \cite{sherson_results_2015}. In particular, the author provides an in-depth explanation of how the denominator in equation \cref{VLTimp3} affects singularities and reconstruction quality.

For readers wishing to explore these topics further, the venerable book by Natterer \cite{natterer_mathematics_1986} is still a comprehensive reference on the classical theory of the Radon transform and filtered back-projection. The recent monograph by Ambartsoumian \cite{ambartsoumian_generalized_2023} provides a thorough treatment of broken-ray, V-line, star, and cone transforms. For a more introductory treatment, the book by Feeman \cite{feeman_mathematics_2010} is accessible to readers with a minimal background in analysis.

\section{Summary}

The attenuation coefficient is the primary physical quantity of interest in many tomography techniques, which are modeled mathematically using various integral transforms.

Beginning with the divergent beam transform, we introduced the Fourier transform as a versatile tool for deriving inversion formulas, exploiting the key properties that differentiation corresponds to multiplication by a frequency variable and integration corresponds to division. We then extended these ideas to the V-line transform, which arises in models of single-scattering tomography. Fixing the directions $\mathbf{u}_1$ and $\mathbf{u}_2$ made the Fourier transform directly applicable. We then derived the explicit inversion formula \cref{VLTinv2} by working algebraically in the Fourier domain and translating the result back into a sequence of differential and integral operators.

These results illustrate the power of the Fourier transform in the inversion of integral operators, suggesting a systematic approach to inverting a broad class of integral transforms that arise in tomography. The methods presented serve as a foundation for more complex imaging problems, including attenuated and limited-data variants. In practice, the Fast Fourier Transform makes these inversion procedures computationally efficient.

\appendix
\section{An Exponential Integral}\label{appendix}

In the main text, we encountered the integral (equation \cref{appendix_eq}):

\begin{equation}\label{keyint}
    \int_0^\infty e^{\imath2\pi t s} \, dt = \frac{-1}{\imath 2 \pi s} + \frac{1}{2}\delta(s),
\end{equation}

where $s = \mathbf{u} \cdot \xi \in \mathbb{R}$ is a real-valued parameter. This integral does not converge in
the classical sense, since the integrand oscillates without decay as $t \to \infty$. This appendix includes a detailed computation of this integral, and requires techniques from distribution theory and real analysis.

\subsection*{Regularization}
Since the integrand in \cref{keyint} does not decay as $t \to \infty$, this integral must be understood in the
distributional sense \cite{strichartz_guide_2003}. A standard technique is to multiply by a convergence
factor and pass to the limit (see \cite[Ch.~8]{folland1992fourier} and \cite[Ch.~9]{rudin_real_1987}). We regularize the integrand by
multiplying by $e^{-\varepsilon t}$ for $\varepsilon > 0$ and take the limit as $\varepsilon \to 0^+$:
\begin{equation}
\int_0^\infty e^{\imath2\pi t s} \, dt
= \lim_{\varepsilon \to 0^+} \int_0^{\infty} e^{-\varepsilon t} e^{\imath 2\pi s t}\, dt
= \lim_{\varepsilon \to 0^+} \int_0^{\infty} e^{-(\varepsilon - \imath 2\pi s)t}\, dt.
\end{equation}

\subsection*{Evaluating the Regularized Integral}
For $\varepsilon > 0$, the exponential $e^{-(\varepsilon - \imath 2\pi s)t}$ decays as $t \to \infty$, so the integral converges and may be evaluated directly:
\begin{equation}
\int_0^{\infty} e^{-(\varepsilon - \imath 2\pi s)t}\, dt
= \left[\frac{e^{-(\varepsilon - \imath 2\pi s)t}}{-(\varepsilon - \imath 2\pi s)}\right]_0^\infty
= \frac{1}{\varepsilon - \imath 2\pi s}.
\end{equation}

\subsection*{Decomposing into Real and Imaginary Parts}
Multiplying the numerator and denominator by the complex conjugate $(\varepsilon + \imath 2\pi s)$, one obtains:
\begin{equation}
\frac{1}{\varepsilon - \imath 2\pi s}
= \frac{\varepsilon + \imath 2\pi s}{\varepsilon^2 + (2\pi s)^2}
= \frac{\varepsilon}{\varepsilon^2 + (2\pi s)^2}
+ \imath\frac{2\pi s}{\varepsilon^2 + (2\pi s)^2}.
\end{equation}
We now pass to the limit $\varepsilon \to 0^+$ in each part separately.

\subsection*{Evaluating the Limits}
The family of functions $\varepsilon \mapsto \frac{1}{\pi}\frac{\varepsilon}{\varepsilon^2+x^2}$
is a classical \textit{approximation to the identity} (also called the Poisson kernel), and
converges in the distributional sense to the Dirac delta distribution \cite{folland1992fourier, strichartz_guide_2003}, defined by the relation:
\[
\int_{-\infty}^\infty g(x)\delta(x)\,dx = g(0).
\]
That is, we use the result:
\begin{equation}
\lim_{\varepsilon\to 0^+}\frac{1}{\pi}\frac{\varepsilon}{\varepsilon^2 + x^2} = \delta(x).
\end{equation}
 and apply this with $x = 2\pi s$:
 
\begin{equation}\label{s1}
\lim_{\varepsilon \to 0^+} \frac{\varepsilon}{\varepsilon^2 + (2\pi s)^2}
= \pi\delta(2\pi s)
= \frac{1}{2}\,\delta(s),
\end{equation}

where in the last step we used the scaling property $\delta(2\pi s) = \frac{1}{2\pi}\delta(s)$. Next, for $s \neq 0$, it holds as $\varepsilon \to 0^+$ that:
\begin{equation}\label{s2}
\imath\frac{2\pi s}{\varepsilon^2 + (2\pi s)^2} \to \frac{\imath}{2\pi s} = \frac{-1}{\imath\, 2\pi s}.
\end{equation}

\subsection*{Result}
Combining the real and imaginary parts \cref{s1} and \cref{s2} and substituting $s = \mathbf{u} \cdot \xi$ recovers the formula used in the main text:
\[
    \int_0^\infty e^{\imath2\pi t(\mathbf{u} \cdot \xi)} \,dt
      = \frac{-1}{\imath 2 \pi (\mathbf{u} \cdot \xi)} + \frac{1}{2}\delta(\mathbf{u} \cdot \xi),
\]

where both terms are understood in the sense of distributions.

\bibliographystyle{siamplain}


\end{document}